# Giant Planar Hall Effect in Dirac Semimetal ZrTe$_{5-\delta}$

P. Li[†], C.H. Zhang[†], J.W. Zhang, Y. Wen, and X.X. Zhang[*]

*King Abdullah University of Science and Technology (KAUST), Physical Science and Engineering Division (PSE), Thuwal 23955-6900, Saudi Arabia*

\* xixiang.zhang@kaust.edu.sa

Recently, giant planar Hall effect originating from chiral anomaly has been predicted in nonmagnetic Dirac/Weyl semimetals. ZrTe$_5$ is considered to be an intriguing Dirac semimetal at the boundary of weak topological insulators and strong topological insulators, though this claim still remains controversial. Here, we report the observation in ZrTe$_{5-\delta}$ of the giant planar Hall resistivity that shows two different magnetic-field dependences as predicted by theory and a maximum at the Lifshitz transition temperature. We found that the giant planar Hall resistivity fades out with decreasing the thickness of ZrTe$_{5-\delta}$ nanoplates, which may be ascribed to the vanishing of the 3D nature of the samples. In addition, we have observed a nontrivial Berry phase, chiral-anomaly-induced negative longitudinal magnetoresistance, and a giant in-plane anisotropic magnetoresistance in these ZrTe$_{5-\delta}$ nanoplates. All the experimental observations demonstrated coherently that ZrTe$_{5-\delta}$ is a Dirac semimetal.

Dirac and Weyl semimetals as new type quantum materials have drawn tremendous attention recently for the novel physics and potential applications.[1-8] One of the most intriguing magnetotransport properties in these materials is the chiral-anomaly-induced negative magnetoresistance (NMR).[5,9,10] Although NMR is considered a signature of Dirac or Weyl semimetals, several extrinsic factors, such as current jetting and conductance fluctuation, may also lead to NMR.[11-14] Very recently, giant planar Hall Effect (PHE) was predicted theoretically to appear in Dirac and Weyl semimetals,[15,16] where PHE refers to the transverse voltage when a magnetic field is applied in in-plane with the electrical current. PHE is a weak magnetic field effect and usually observed in ferromagnetic materials where it originates from the spin-orbit coupling.[17-19] Different from in ferromagnetic material, the PHE in topological Dirac and Weyl semimetals was theoretically demonstrated to originate from nontrivial Berry phase and chiral anomaly[15,16,20-24]. The angular dependence of PHE resistivity and longitudinal anisotropic magnetoresistance in Dirac and Weyl semimetals can be described as[15,16]

$$\rho_{xy} = -\Delta\rho^{chiral} \sin\varphi \cos\varphi, \qquad (1)$$

$$\rho_{xx} = \rho_\perp - \Delta\rho^{chiral} \cos^2\varphi, \qquad (2)$$

where $\Delta\rho_{chiral} = \rho_\perp - \rho_{//}$ is resistivity anisotropy induced by chiral anomaly and berry phase, $\rho_\perp$ and $\rho_{//}$ are the resistivity for a magnetic field applied transversely and parallel to the current in the current plane, respectively.

The monolayer $ZrTe_5$ was initially predicted and demonstrated to be a 2D quantum spin Hall insulator.[25-27] Although Angle-resolved photoemission spectroscopy (ARPES) experiments showed directly that a bulk $ZrTe_5$ is a weak topological insulator (TI),[28] it was also demonstrated that bulk $ZrTe_5$ is a 3D Dirac semimetal by the observation of chiral anomaly in magnetotransport[29] and nontrivial Berry phase,[30,31] and by magneto-infrared spectroscopy.[32] More interestingly, it has been shown that the Dirac semimetal states will appear in $ZrTe_5$ at the boundary of strong and weak TIs.[33-35] Therefore, the nature of layered transition metal chalcogenide $ZrTe_5$ is still under hot debate and more evidence is needed to uncover its Dirac semimetal state.

In this letter, we reported the observation of chiral-anomaly-induced planar Hall Effect in $ZrTe_{5-\delta}$ nanoplates. The nontrivial Berry phase (~ $\pi$) in $ZrTe_{5-\delta}$ was obtained from the Shubnikov de Haas (SdH) oscillation data. The chiral-anomaly-induced giant planar Hall resistivity reached the maximum value of 254.0 $\mu\Omega$ cm (14 T) near the resistivity anomaly temperature ~150 K, which strongly supports the existence of Dirac point in $ZrTe_{5-\delta}$.

$ZrTe_{5-\delta}$ single crystals grown by iodine assisted vapor transport technique were obtained from Hqgraphene Company. The thin plates of $ZrTe_{5-\delta}$ were exfoliated mechanically onto the $SiO_2$(280 nm)/Si substrates. The electrodes were patterned by standard e-beam lithography and deposited by e-beam evaporation. The magneto-transport measurements were carried out on a physical property measurement system (Dynacool system, Quantum Design Inc.) by standard lock-in methods.[36] The crystal structure of $ZrTe_{5-\delta}$ was imaged using a monochromated Cs-corrected high-resolution scanning transmission electron microscopy. The ionic gating on $ZrTe_{5-\delta}$ devices was carried out using $LiClO_4$+PEO+methanol as the solid electrolyte.[37,38]

Figure 1(a) shows the optical image of a typical exfoliated $ZrTe_{5-\delta}$ micro-ribbon device and measurement configuration of PHE ($\rho_{xy}$) and AMR ($\rho_{xx}$).[36] The high angle area dark field image of $ZrTe_{5-\delta}$ (*ab* plane) is shown in Fig. 1(b). The unit cell of $ZrTe_{5-\delta}$ in *a-b* plane is indicated by the white dashed rectangle, in which Zr and Te atoms are highlighted with green and purple dots for clarity. The lattice constant along *b* is identified to be ~14.5 Å. According to the previous report, ZrTe$_5$ will transform from strong TIs to weak TIs with increasing the lattice constant $b$ larger than 14.46 Å, whereas the Dirac semimetal states will appear at the boundary of the transformation.[33,34] Therefore, the lattice constant of our sample supports the existence of Dirac semimetal state.

Figure 1(c) shows the typical temperature dependence of the longitudinal resistance of ZrTe$_{5-\delta}$ under magnetic fields of zero and 14 T($I//a$, $B//b$). A pronounced resistance anomaly, a metal-insulator transition (MIT), appeared at ~135 K in the zero-field curve, which agrees well with previous results.[38,39] Notably, the transition temperature is slightly increased to ~150 K as the magnetic field increased to 14 T. This MIT transition was ascribed to the consequence of the Lifshitz transition[40] that was accompanied with the change of the dominated carrier type, from $n$-type (low temperatures) to $p$-type (high temperatures).[28] It is reported recently that ZrTe$_5$ single crystals grown by chemical vapor transport are of Te deficiency (ZrTe$_{5-\delta}$) and that the MIT near 130 K should be correlated to the bipolar conduction.[41] The metallic behavior at low temperatures should be the critical requirement for a Diarc/Weyl semimetal, which was observed in our samples. The single crystals grown by flux method (Flux) have a ZrTe$_5$ stoichiometry with a MIT temperature below 5 K. More importantly, it is demonstrated that the stoichiometric ZrTe$_5$ is a narrow gap semiconductor instead of a semimetal.[41] Elemental analysis results show that our samples is ZrTe$_{5-\delta}$ ($\delta$=0.22, ZrTe$_{4.78}$) with Te vacancies, consistent with that reported.[41] Not surprisingly, we also observed a sign change in Hall resistance near 135 K accompanied by the change of dominated carrier from $n$-type to $p$-type, as shown in Fig. 1(d) and its inset.[36] The anomalous Hall effect at low magnetic fields in Fig. 1(d) was observed in previous reports,[38,40,42] which were interpreted within the frameworkof Berry curvature generated by Weyl node in ZrTe$_5$.[42]

To explore the physics of the quantum magneto-transport underlying the experimental data (Fig. 2(a)), we calculated $d\rho/dB$ from the transverse magnetoresistance measured with magnetic field applied perpendicular to the current plane ($B//b$-axis) and in the current plane to reveal the SdH oscillations. The SdH oscillations are clearly seen in the curves of $d\rho/dB$ as a function of $B^{-1}$ for different temperatures (Fig. 2(b)). A small shoulder at $B=5.3$ T, indicated by the arrow, can be ascribed to the effect of spin splitting.[30] To extract the Berry phase, we plotted the Landau fan diagram in Fig. 2(c). According to the Lifshitz-Onsager quantization rule, $B_F/B = n - \gamma + \delta$, where $n$, $B_F$ and $\gamma$ are respectively the Landau index, Oscillation frequency and the Onsager phase factor, $\gamma = 1/2 - \phi_B/2\pi$. $\delta$ is an additional phase shift whose value is within the range of $\pm 1/8$ and depends on the degree of the dimensionality of the Fermi surfaces. The Berry phase $\phi_B$ can then be obtained from the intercept of Landau fan curve in Fig. 2(c). To avoid the effect of spin splitting at high fields on the determination of Berry phase,[6] we only linearly fitted the Landau index with $n>3$ as a function of $B^{-1}$. From the intercept obtained from the linear fitting, we find that a nontrivial Berry phase $\phi_B = 1.07\pi \pm 0.02\pi$ [36] and that this value agrees very well with that reported previously.

This nontrivial Berry phase motivated us to further explore the chiral-anomaly-induced NMR and other exotic transport properties in our ZrTe$_{5-\delta}$ samples. Figure 2(d) shows the longitudinal MR at different temperatures measured with $B//I//a$-axis in the current plane. In addition to the pronounced SdH oscillations, the negative MR was clearly seen at low temperatures ($T<40$ K) under strong magnetic fields. Based on the geometry of our devices, we can exclude the contribution of current jetting effect to the observed NMR.[11,12,36]

The nontrivial Berry phase and NMR support strongly the existence of a topological semimetal state in ZrTe$_{5-\delta}$ and suggest that a giant PHE could be observed in our sample ZrTe$_{5-\delta}$ as a signature of topological semimetal state.[15,16,20-23] Figures 3(a)-3(c) show the angular dependence of the planar Hall resistance $R_{xy}^{planar}$ of ZrTe$_{5-\delta}$ measured at different temperatures, where the sample is rotating in *ac* plane. Apparently, the data of $\rho_{xy}^{planar}$ measured at 2 K and 200 K and under both fields do not follow the sin2$\varphi$ ($\varphi$: angle between *B* and *I* in *ac* plane) dependence (Eq. (1)) thoroughly, but the data measured at 150 K indeed can be described by a sin2$\varphi$ angular dependence. Another interesting feature is that the main peak in $\rho_{xy}^{planar} - \varphi$ curves for *B*=14 T shifts gradually from ~310° below 150 K to ~130° above 150 K. Based on the data in Fig. 1d, we understand that the dominant carrier changes from *n*-type to *p*-type at about 135 K. This asymmetric angular dependence of $\rho_{xy}^{planar}$ could be due to the contribution of a normal Hall effect that arose from on the perpendicular component of the applied magnetic field. This perpendicular component could be easily caused by the small misalignment of the sample surface (*ac* plane) respect to the magnetic field, as shown in Fig. 3d.

To separate the normal Hall contribution from $\rho_{xy}^{planar}$, we calculated the average of the measured $\rho_{xy}^{planar} - \varphi$ under the -14 T and 14 T, as the green curves shown in Figs. 3(a)-3(c). Typically, we found that $\rho_{xy}^{planar}$ becomes much more symmetric (Fig. 3(e)) and shows two-fold feature over 360°. However, the data cannot be directly described by Eq. (1) due to a resistivity shift of 150 μΩ cm away from zero. This resistivity shift should arise from the longitudinal misalignment during the fabrication of Hall bar device. This small longitudinal resistivity contribution can be subtracted by fitting the averaged data to Eq.(3):

$$\rho_{xy}^{planar} = -\Delta\rho_{xy}^{chiral} \sin\varphi\cos\varphi + a\Delta\rho_{xy}^{chiral} \cos^2\varphi + b. \tag{3}$$

The first term is the intrinsic PHE originating from chiral anomaly; the second and third terms are respectively the in-plane AMR and longitudinal resistance offset caused by the misalignment of Hall bar. However, compared to the standard longitudinal anisotropic magnetoresistance observed in these devices (Figs. 4(a)), the angular dependent longitudinal misalignment-resistance is quite small, evidenced by the small resistance shift in Hall effect (Fig. 1(d)). After taking the longitudinal Hall bar misalignment-resistance $b$ into account, the $\rho_{xy}^{planar}$ can be well fitted by Eq. (3), as the red line shown in Fig. 3(e). To exclude the artifacts in the analysis of the raw data, we also fitted measured $\rho_{xy}^{planar}$ directly with three contributions, two-fold PHE, one-fold Hall effect, and resistance offset due to the misalignment of Hall bar, as shown in Fig. 3(f). We found that the amplitude of the intrinsic $\rho_{xy}^{planar}$ obtained from above two fitting strategies are identical, 67.00±0.60 μΩ cm in Fig. 3(e) and 66.60±0.60 μΩ cm in Fig. 3(f), which suggests that both methods result to nearly the same intrinsic $\rho_{xy}^{planar}$.

To further confirm our argument, we measured the longitudinal AMR ($\mathrm{AMR} = (R_\varphi - R_\perp)/R_\perp$) at different temperatures as shown in Fig. 4(a). For comparison, the angular dependent intrinsic planar Hall resistivity $\rho_{xy}^{chiral}$ measured under different magnetic fields are shown in Fig. 4(b). The exact 45° phase difference between the angular dependent two-fold AMR and intrinsic PHE agree well with the theoretical predictions and Eqs. (1) and (2).[15,16,36] Generally, the anisotropic magnetoresistance in conventional magnetic materials is quite small and is caused by the spin orbit coupling.[17] The giant AMR in this study reaches as high as -43% at 2 K and 14 T, which is believed to be closely related to the giant PHE induced by chiral anomaly.[21,22]. It should also be noted that $\rho_{xy}^{chiral}$ increases monotonically to 254.0±2.0 μΩ cm as the magnetic field is increased to 14 T. This giant PHE observed in our ZrTe$_{5-\delta}$ device is about four orders of magnitude higher than that observed in conventional ferromagnetic metals.[19]

To gain a deeper insight into the physical mechanism underlying the giant PHE, we plotted typical $\rho_{xy}^{chiral} - B$ curve for $t = 67$ nm thick device in Fig. 4(c).[36] Apparently, $\rho_{xy}^{chiral}$ does not follow a simple linear or quadratic dependence on the magnetic field $B$ as the MR and Hall effect observed in conventional ferromagnetic materials.[17,18] Instead, it shows roughly two different field-dependences: for low fields ($B<3$ T), $\rho_{xy}^{chiral}$ depends on $B^2$; with increasing field further, $B>6$ T, $\rho_{xy}^{chiral}$ shows a weak tendency of saturation. Actually, it has been theoretically predicted that the chiral-anomaly-induced planar-Hall-effect can be divided into different magnetic field regions:[15]

$$L_a \gg L_c, \quad \rho_{xy}^{chiral} \propto \left(\frac{L_c}{L_a}\right)^2 \propto B^2 \text{ ; weak magnetic field region} \quad (4)$$

$$L_a \ll L_c, \quad \rho_{xy}^{chiral} \propto \frac{1}{\sigma}(1-\frac{2L_a}{L_x}) \text{ for } L_a < L_x < \frac{L_c^2}{L_a} \text{ ; strong magnetic field region} \quad (5)$$

$$L_a \ll L_c, \quad \rho_{xy}^{chiral} \propto \frac{1}{\sigma}(1-\frac{L_a^2}{L_c^2}) \text{ for } L_x > \frac{L_c^2}{L_a} \text{ ; strong magnetic field region} \quad (6)$$

where $\sigma$, $L_x$, $L_a$, and $L_c$ are conductivity, sample length, magnetic length ($L_a \propto B^{-1}$), and chiral charge diffusion length.[15] The weak tendency of saturation at strong magnetic field will follow a $-B^{-1}$(Eq. (5)) or $-B^{-2}$(Eq. (6)) relation, depending on the sample length $L_x$.[15] The red line in Fig. 4(c) gives the fitting by Eq. (5) shows the typical weak tendency of saturation at high magnetic fields. The consistency between our observation and theory is the strong evidence of the existence of topological state in our samples.

To understand the behavior of $\rho_{xy}^{chiral}$ in more detail, we investigated the dependence of the giant planar Hall effect on the thickness of ZrTe$_{5-\delta}$ nanoplates.[36] Figure 4(d) shows the amplitude of the planar-Hall-resistivity with various $ZrTe_5$ nanoplate thicknesses as a function of the temperature under a magnetic field of 14 T, in which the values of planar-Hall-resistivity are taken from the angular dependence of $\rho_{xy}^{chiral}$ measured at different temperatures. Interestingly, a maximum intrinsic planar Hall resistivity $\rho_{xy}^{chiral}$ appear in most of samples except for two ultrathin $ZrTe_{5-\delta}$ devices ($t$=25 and 30 nm). Particularly important, very strong and well defined peaks occur in the temperature range of 110~150 K in the vicinity of the corresponding MIT temperatures. This peak behavior of $\rho_{xy}^{chiral}$ ($T$) is different from the monotonic decrease of $\rho_{xy}^{chiral}$ with increasing temperature observed in some Dirac/Weyl semimetals.[20-23] An ARPES study showed that the Fermi level in $ZrTe_{5-\delta}$ will sweep from electron-like band to hole-like band across MIT temperature,[28] though no definite evidence can be found to support the existence of Dirac cone in the ARPES spectra.[28] However, our observation of giant PHE and its interesting behavior in Fig. 4(d) will be a piece of strong evidence to support the existence of Dirac fermions in $ZrTe_{5-\delta}$, since the giant planar-Hall-effect is caused by the chiral anomaly and berry phase of carriers near Dirac point.[15,16] When the Fermi level in $ZrTe_{5-\delta}$ is sweeping across Dirac point near MIT temperature, the chiral-anomaly-induced intrinsic $\rho_{xy}^{chiral}$ will be enhanced (Fig. 4(d)). Nevertheless, in other well-known Dirac/Weyl semimetals, such as $Cd_3As_2$, GdPtBi, and $WTe_2$,[20-23] their Dirac point will deviate from Fermi level further as the temperature increases.[43,44] Therefore, a monotonic decrease of $\rho_{xy}^{chiral}$ with increasing temperature is observed in these Dirac/Weyl semimetals. Another striking feature in Fig. 4(d) is the monotonic decrease of planar Hall resistivity as the decreasing of device thickness. This is a typical characteristic and further indication that $ZrTe_{5-\delta}$ is a 3D Dirac/Weyl semimetal.

To further consolidate our argument, we investigated the effect of ionic gating on planar Hall effect in ZrTe$_{5-\delta}$. Given to the relative high carrier density in ZrTe$_{5-\delta}$, it is almost impossible to tune the carrier density in thick devices (e.g. $t$=174 nm).[36] We indeed observed the shift of MIT temperature from 120 K ($V_g$=0 V) to 340 K ($V_g$=3.0 V) under the ionic gating in thin device ($t$=30 nm), in agreement with previous gating experiment.[38] Nevertheless, limited by the relative weak planar Hall effect in thin ZrTe$_{5-\delta}$, it is hard to investigate the gating effect on planar Hall effect.[36] Therefore, we carefully examined the gating effect on a device with moderate thickness ($t$=58 nm), as displayed in Fig. 4(e). By applying $V_g$=1.5 V, the transition temperature in $\rho-T$ curve shifted from 118 K down to 106 K. When the gating voltage was increased to 2.5 V, the transition temperature was increased to 146 K. This nonmonotonic dependence agrees well with previous report.[38] More important, the maximum in the temperature dependent planar Hall resistivity occurs at the same temperature where the MIT transition appears in the corresponding $\rho-T$ curve, which strongly indicates that the planar Hall effect in ZrTe$_{5-\delta}$ is a direct consequence of chiral anomaly related to Dirac/Weyl point.

To conclude, we have observed a giant planar Hall Effect in ZrTe$_{5-\delta}$, accompanied with a nontrivial Berry phase, negative longitudinal magnetoresistance, and a giant anisotropicmagnetoresistance, which demonstrated fully that ZrTe$_{5-\delta}$ with the MIT temperature near 130 K is indeed a Dirac semimetal.

## Acknowledgement

Peng Li and Chenhui Zhang contribute equally to this work. We thank Prof. Feng Liu in University of Utahfor very helpful discussions. The research reported in this publication was supported by King Abdullah University of Science and Technology (KAUST). P. Li acknowledges the financial support of CRF-2015-SENSORS-2709 (KAUST).

# Figure Captions

**FIG. 1 (Color online)** (a) Optical image of ZrTe$_{5-\delta}$ devices. The scale bar is 30 μm. (b) HAADF image of ZrTe$_{5-\delta}$ sample in *ab* plane. The dashed rectangle gives the unit cell of ZrTe$_{5-\delta}$ in *ab* plane. The green and purple dots represent Zr and Te atoms, respectively. The scale bar is 1 nm. (c) Temperature dependence of resistivity of ZrTe$_{5-\delta}$ under different magnetic fields. (d) Hall resistivity of ZrTe$_{5-\delta}$ as a function of magnetic field at selected temperatures. The inset gives $\rho_{xy} - T$ curves (*B*=5 T).

**FIG. 2 (Color online)** (a) Magnetic field dependent magnetoresistance ratio of ZrTe$_{5-\delta}$ with $B//b$. (b) SdH oscillation amplitude as function of $B^{-1}$ at different temperatures. The arrow indicates the spin splitting at high fields. (c) Landau fan diagram of ZrTe$_{5-\delta}$. (d) Longitudinal magnetoresistance with $B//a$ at different temperatures.

**FIG. 3 (Color online)** Measured planar Hall resistivity $\rho_{xy}^{\text{planar}}$ of ZrTe$_{5-\delta}$ ($t$=200 nm) under the opposite magnetic fields (14 T and -14 T), (a) $T$=2 K; (b) $T$=150 K; (c) $T$=200 K. (d) The schematic gives the measurement configuration with the misalignment that includes out-of-plane magnetic field component. The yellow circle gives the ideal rotation of magnetic field during the measurement of PHE. The blue circle represents the real ration of magnetic field with misalignment, which will give the contribution of Hall effect in measured $\rho_{xy}^{\text{planar}}$. (e) Typical fitting of averaged $\rho_{xy}^{\text{planar}}$ and the fitting curves ($T$=200 K). (f) Typical fitting of angular dependent $\rho_{xy}^{\text{planar}}$ with three contributions: intrinsic PHE, Hall effect, and longitudinal resistivity offset.

**FIG. 4 (Color online)** (a) Measured in-plane giant anisotropic magnetoresistance of ZrTe$_{5-\delta}$ at different temperatures ($B$=14 T). (b) Intrinsic planar Hall resistivity $\rho_{xy}^{chiral}$ as a function of rotation angle at different magnetic fields ($T$=150 K). (c) The magnetic field dependence of chiral-anomaly-induced planar Hall resistivity $\rho_{xy}^{chiral}$. It can be divided into weak and strong magnetic field regions. The red line gives the fitting by Eq. (5) for weak tendency of saturation at strong magnetic fields. The error bar comes from the fitting by Eq. (3). (d) The intrinsic planar Hall resistivity $\rho_{xy}^{chiral}$ as a function of temperature ($B$=14 T) with various ZrTe$_{5-\delta}$ nanoplate thicknesses. (e) The ionic gating effect on planar Hall effect ($t$=58 nm). The maximum temperature in $\rho_{xy}^{chiral} - T$ ($B$=14 T) curves corresponds to the MIT temperatures as the gating voltage increases from 0 V to 2.5 V.

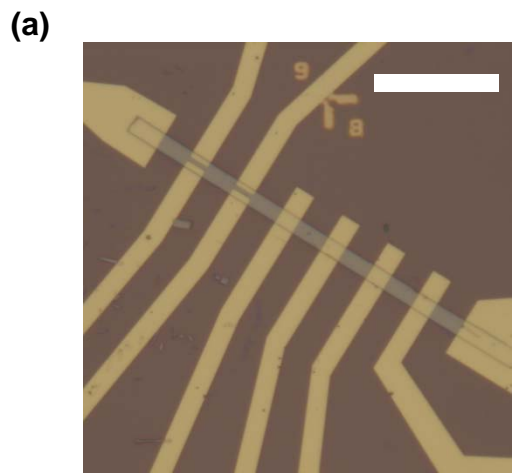
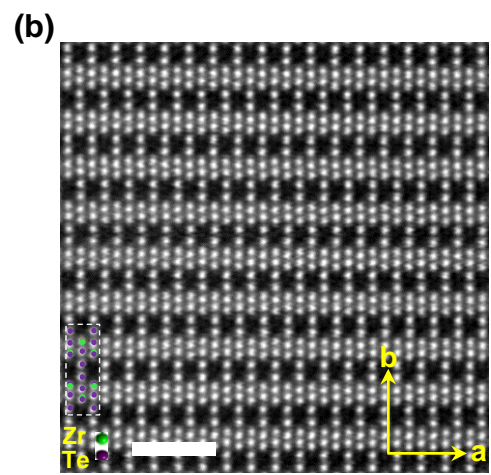
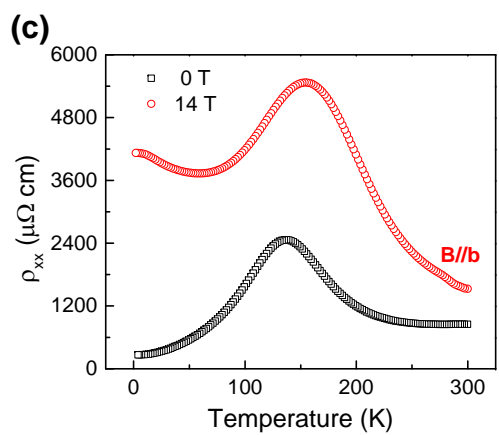
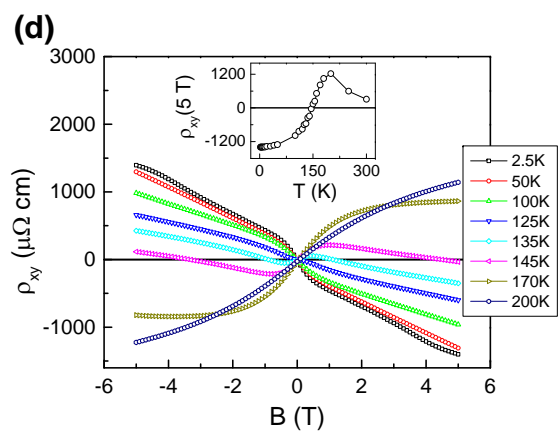

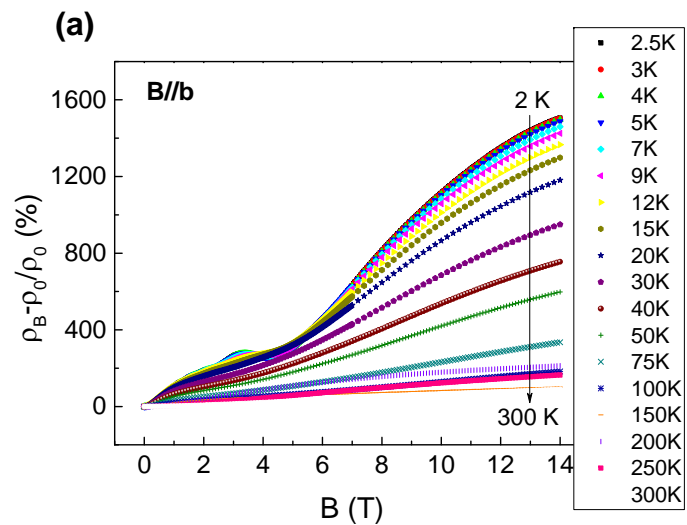
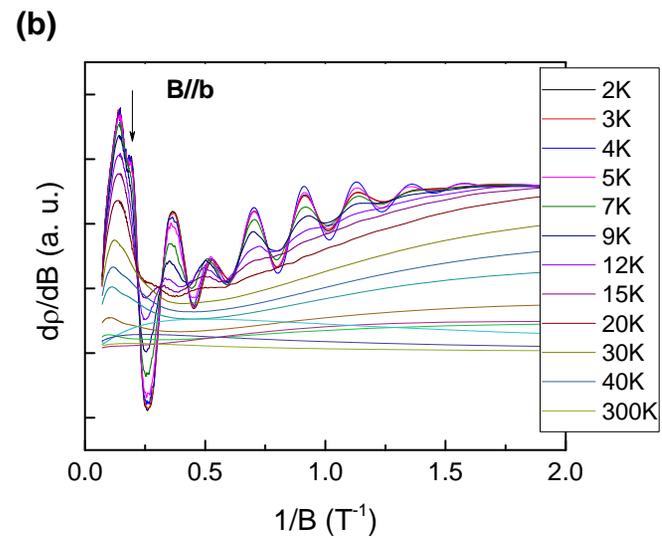
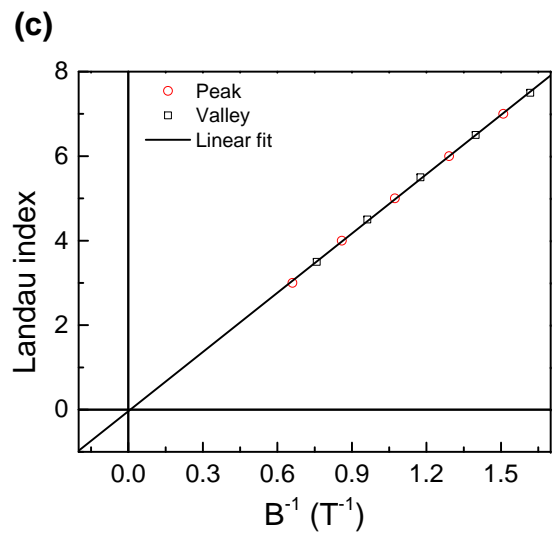
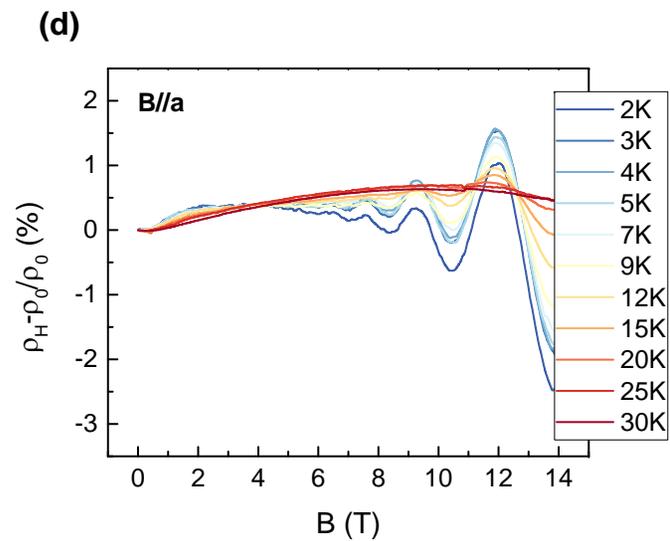

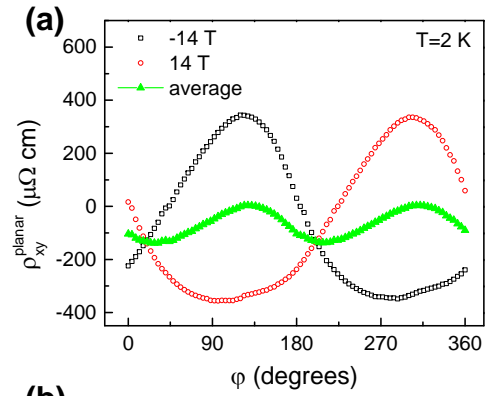
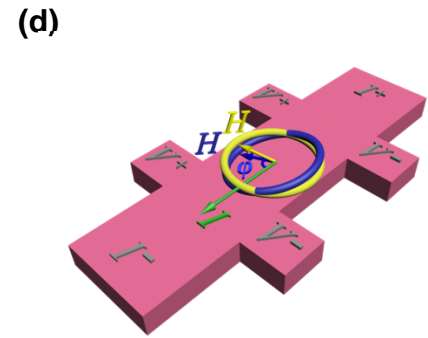
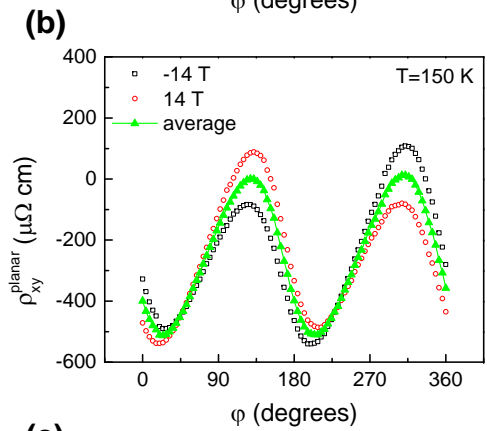
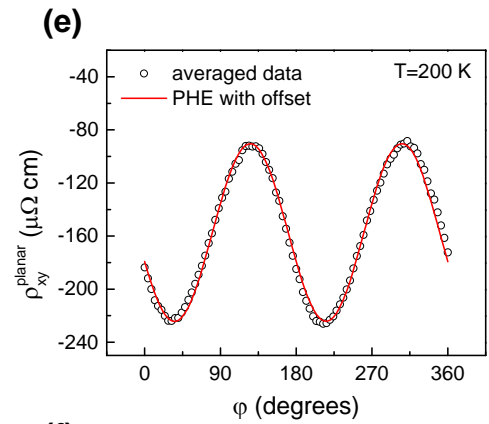
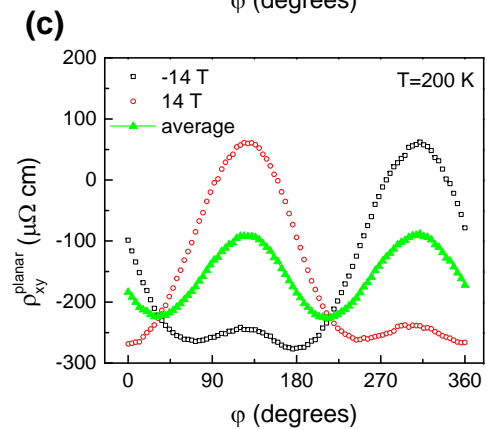
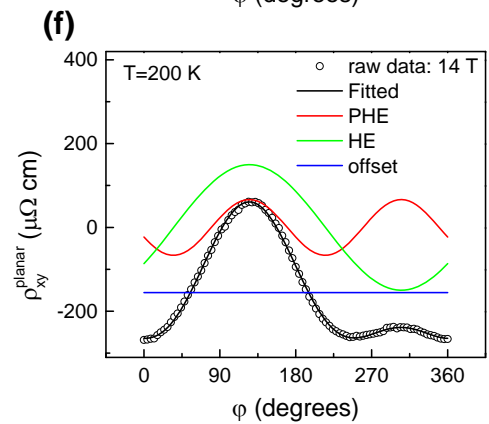

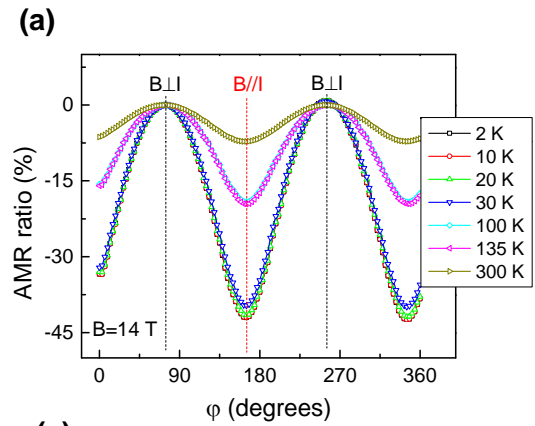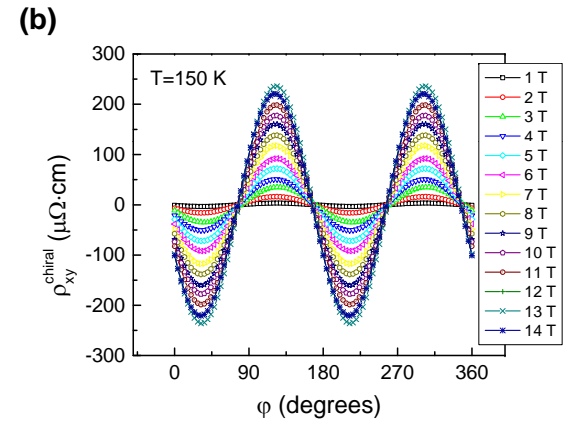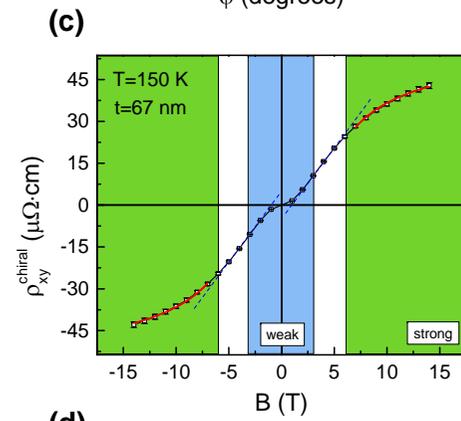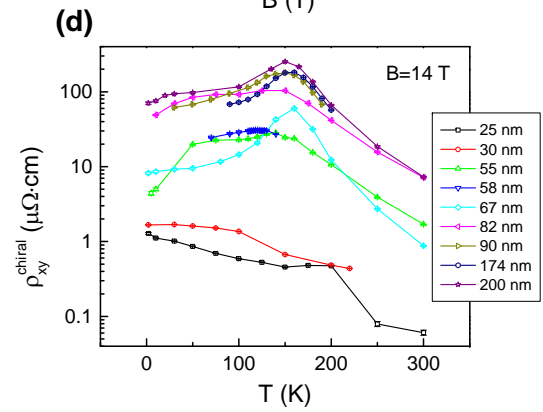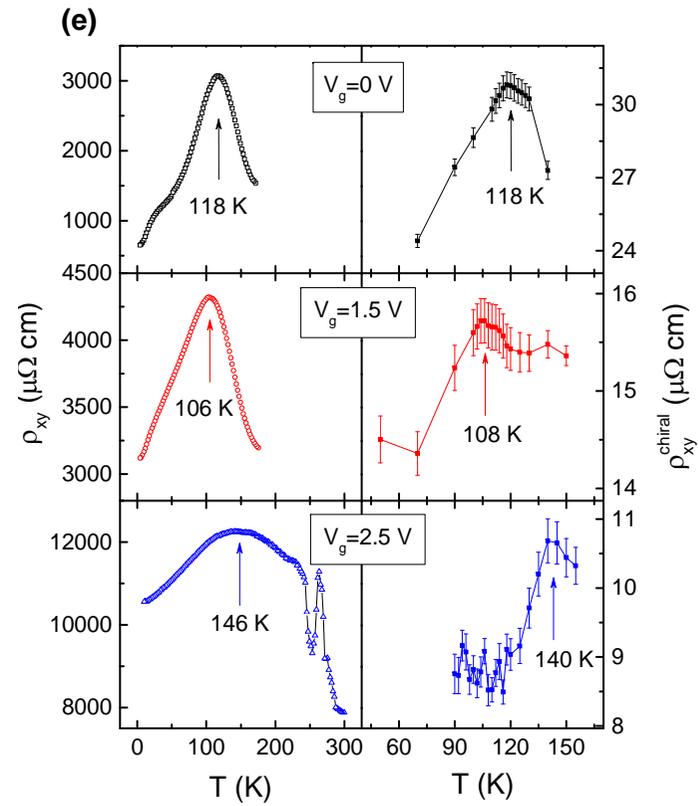

# Supplementary Materials for

## Giant Planar Hall Effect in the Dirac Semimetal ZrTe$_{5-\delta}$

P. Li[†], C.H. Zhang[†], J.W. Zhang, Y. Wen, and X.X. Zhang[*]

*King Abdullah University of Science and Technology (KAUST), Physical Science and Engineering Division (PSE), Thuwal 23955-6900, Saudi Arabia*

*xixiang.zhang@kaust.edu.sa

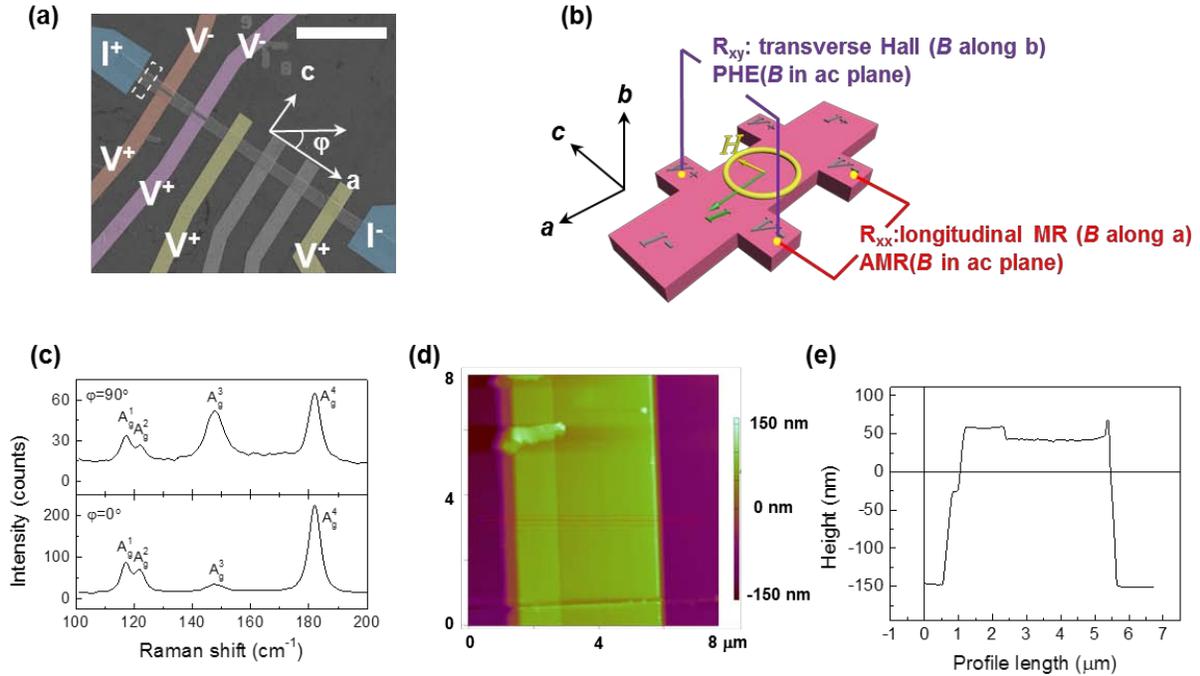

**FIG.1S** (a) Typical scanning electron microscope (SEM) image of ZrTe$_{5-\delta}$ device. The scale bar is 30 μm. Two yellow electrodes are used for the measurement of resistance $R_{xx}$, and two pink or orange electrodes are for Hall resistance (normal Hall effect and planar Hall effect) $R_{xy}$. (b) Measurement configuration of longitudinal magnetoresistance, anisotropic magnetoresistance (AMR), transverse Hall effect (normal Hall), planar Hall effect (PHE) in our ZrTe$_{5-\delta}$ device, (c) Typical Raman spectrum of ZrTe$_{5-\delta}$ along different in-plane crystal axis. The in-plane orientation angle is labeled in (a). All four peaks, especially quite strong anisotropic mode $A_g^3$ are consistent with previous report.[1] (d) Atomic force image of ZrTe$_{5-\delta}$ device in the dashed rectangle region in (a). (e) Line scan profile to determine the thickness of ZrTe$_{5-\delta}$ device ($t \approx 200$ nm).

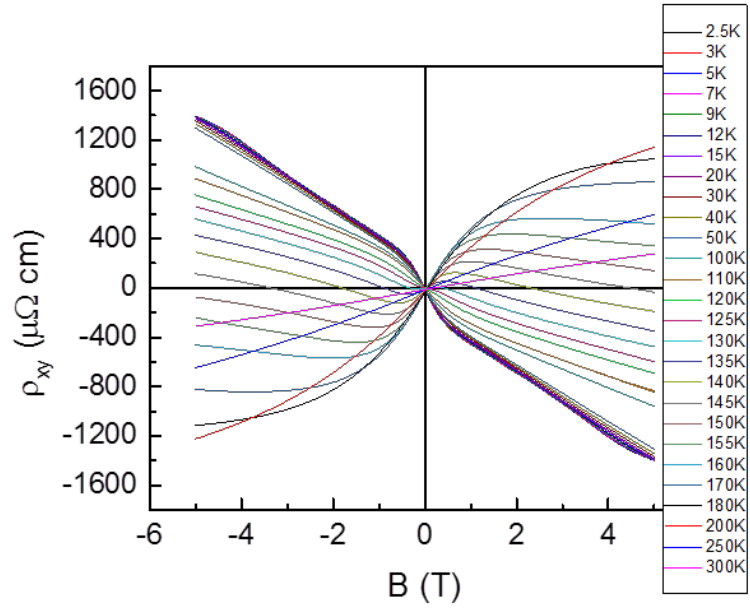

**FIG.2S** Magnetic field dependent Hall resistivity at various temperatures, which is more detailed than that in the inset of Fig. 1(d).

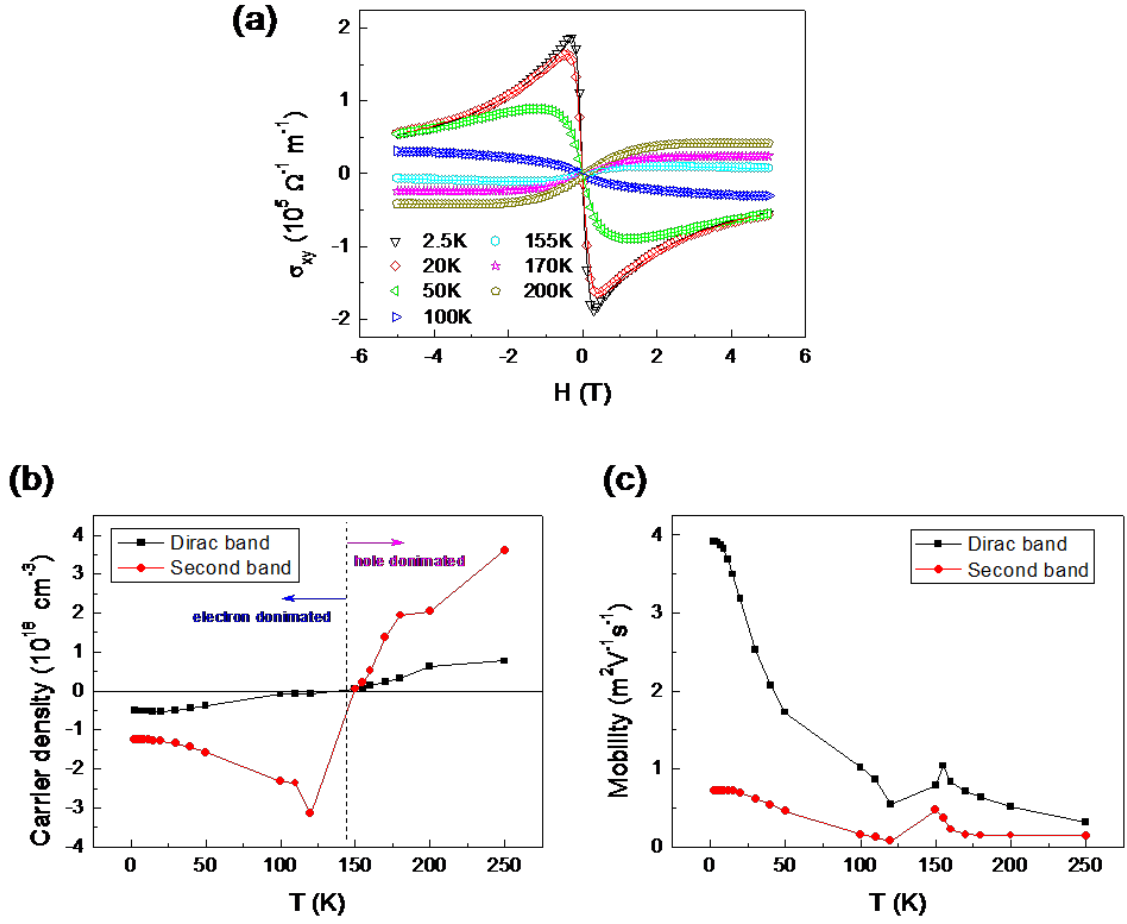

**FIG.3S** (a) Typical magnetic field dependent Hall conductivity at various temperatures, the solid line is the linear fit by two-band model.[2-4] The fitting parameters, including the Dirac band carrier density and second band carrier density and corresponding mobility, are given in (b) and (c) The carrier type change the type near MIT transition temperatures.

We can obtain the carrier density and mobility through two-band model,[2,3]

$$\sigma_{xy} = \frac{n_1 e \mu_1^2 H}{1+\mu_1^2 H^2} + \frac{n_2 e \mu_2^2 H}{1+\mu_2^2 H^2},$$

where $n_1(n_2)$ and $\mu_1(\mu_2)$ are the carrier density and mobility for the Dirac band and the second band. shows the fitted dependence of carrier density and mobility of both bands on temperature.

Obviously, in the metallic and insulating regime ($T$c~ 135 K), the dominated carrier types are *n*-type (negative carrier density) and *p*-type (positive carrier density), respectively. And the mobility of Dirac band is much higher than the second band. Our observation agrees well with previous work.[3] The fitted parameters are also consistent with the physical picture of giant planar Hall effect. When the Fermi level is near the Dirac point, the planar-Hall-effect caused by the chiral anomaly and berry phase of carriers will be maximum.

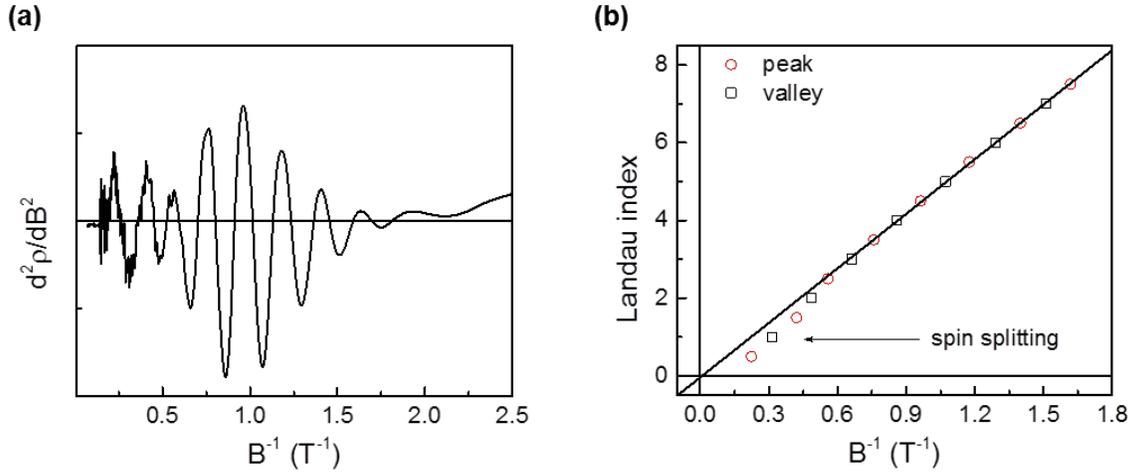

**FIG.4S** (a) Quantum oscillation ($d^2\rho_{xx}/dB^2$, ) as a function of inverse magnetic field with $B//b$ at 2 K. The location of peaks and valleys are obtained for the plot of Landau fan diagram in (b). The deviation of peaks and valley from the linear fitting comes from the spin splitting at high magnetic fields.[5] To obtain the correct Berry phase, we use the data with Landau index $n>3$ to make the linear fit.

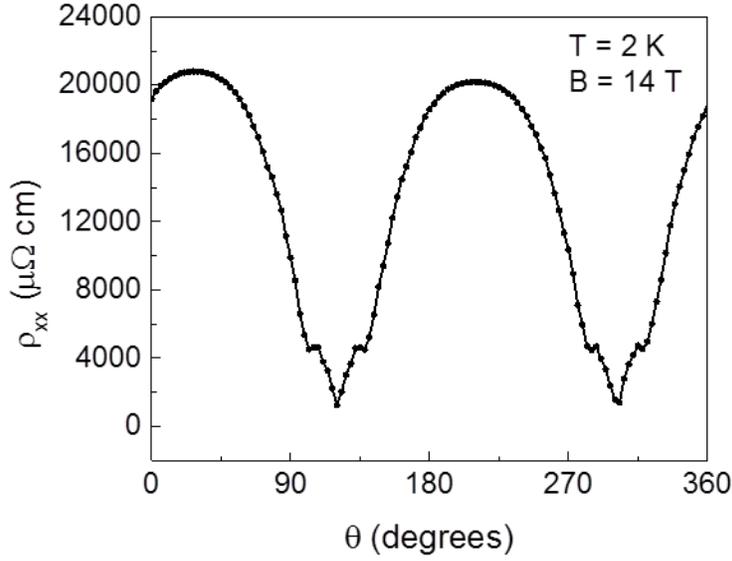

**FIG.5S** Angular dependence of resistivity of ZrTe$_{5-\delta}$ ($t$ =200 nm) with the magnetic field rotated in *ab* plane. $\theta$ is the angle between device plane and vertical magnetic field. The resistivity anisotropy of ZrTe$_{5-\delta}$ under a magnetic field of 14 T is 15.99 ($\rho(I \perp B)/\rho_0$=20796/1300=15.99).

Firstly, the effect of current jetting should be taken into account especially for the point-like contact with sample. Actually, to obtain an Ohmic contact and exclude the effect of residual photoresist on the resistance during nanofabrication, we carefully remove the residual photoresist by Ar plasma before the deposition of electrodes. Moreover, as shown in the microstructure of our sample FIG. 1S(a), all the nanoribbons are well surrounded by the Ti (10 nm)/Au (70 nm), instead of a point-like contact. The quite linear current-voltage curves indicate that it is Ohmic contact.

In the case of current jetting, dips, humps and negative voltage will appear in the angular dependence of longitudinal resistance $R(\theta)$. To confirm the validity of our data, we measured $R(\theta)$, as displayed in FIG. 5S. Although four dips were observed in the $R(\theta)$ curve, four symmetric dips come from the quantum oscillation near 4-5 T in FIG. 2(a). In our observation, we did not observe any negative voltage.

The strong evidence to exclude the current jetting is enough large aspect ratio $l/w$ of the sample length $l$ and width $w$. According to the requirement of homogenous current distribution, the aspect ratio $l/w$ should be larger than $\sqrt{A}$, where $A$ is the resistance anisotropy.[6,7] The resistance anisotropy $A$ at 14 T in FIG. 4S is 15.99 ($\rho(I \perp B)/\rho_0$=20796/1300=15.99), the aspect ratio $l/w$ is, therefore, required to be higher than $\sqrt{15.99} \approx 4$. In our micro-ribbon devices, the width is 4 μm and the length between two voltage probes ($V^+$, $V^-$) is 34 μm with a typical aspect ratio $l/w$ ~8.5, which is larger than the square root of aspect ratio, ~4. Hence, the large aspect ratio $l/w$ in our nanoribbons ensures a homogenous current distribution between two voltage probes.

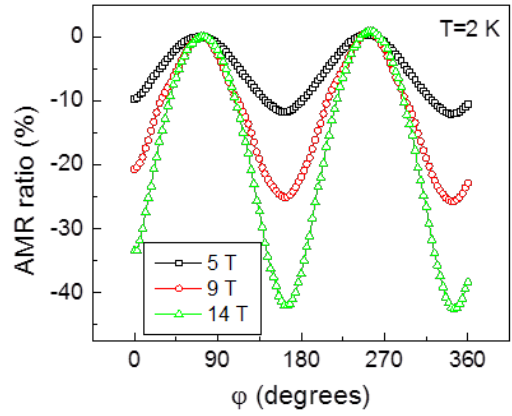

**FIG.6S** In-plane (*ac* plane) magnetoresistance of ZrTe$_{5-\delta}$ under different magnetic fields (*t* =200 nm, *T*=2 K). In the measurement, the current is along *a*-axis and the magnetic field is rotated in *ac* plane.

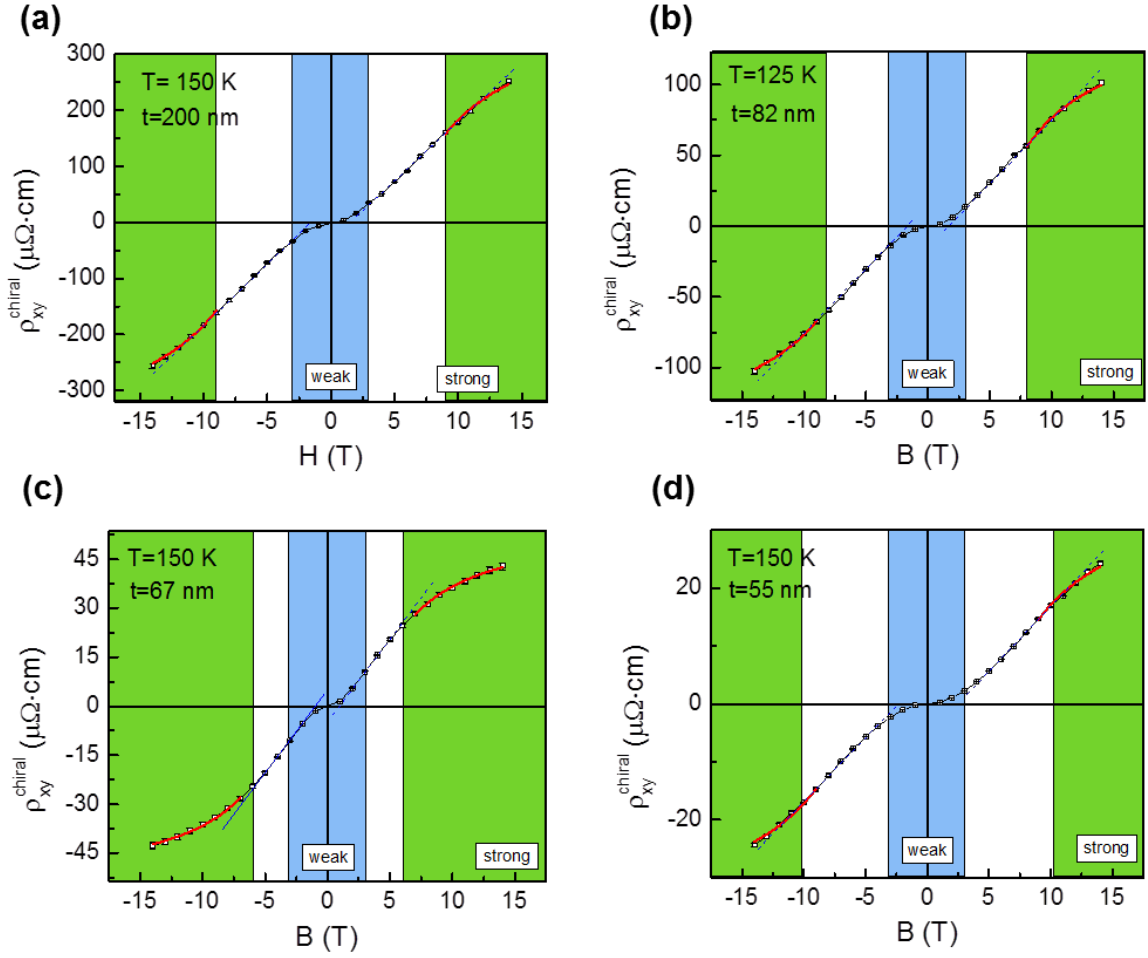

**FIG.7S** Magnetic field dependence of chiral-anomaly-induced planar Hall resistivity $\rho_{xy}^{chiral}$ with different ZrTe$_{5-\delta}$ thicknesses. (a) $t$=200 nm; (b) $t$=82 nm; (c) $t$=67 nm; (d) $t$=55 nm. The weak tendency in saturation of $\rho_{xy}^{chiral}$ at high magnetic fields can be found in all the samples. The red line is the fitting by Eq. (5) in main text.

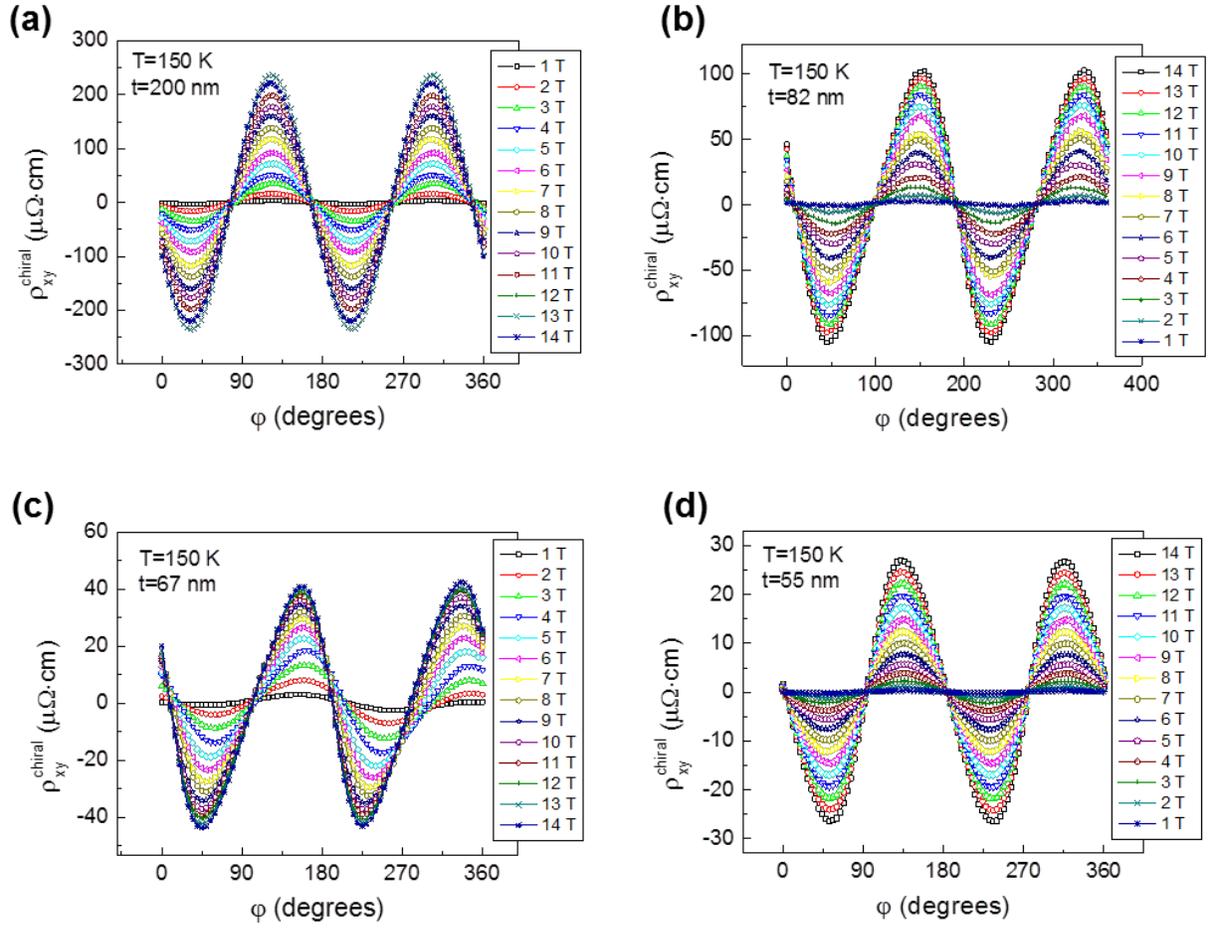

**FIG.8S** Raw data of angular dependent intrinsic planar Hall resistivity under different magnetic fields for devices with various thicknesses (*T*=200 K). (a) *t*=200 nm; (b) *t*=82 nm; (c) *t*=67 nm; (d) *t*=55 nm. The field dependent $\rho_{xy}^{chiral}$ is summarized and plotted in Fig. 7S.

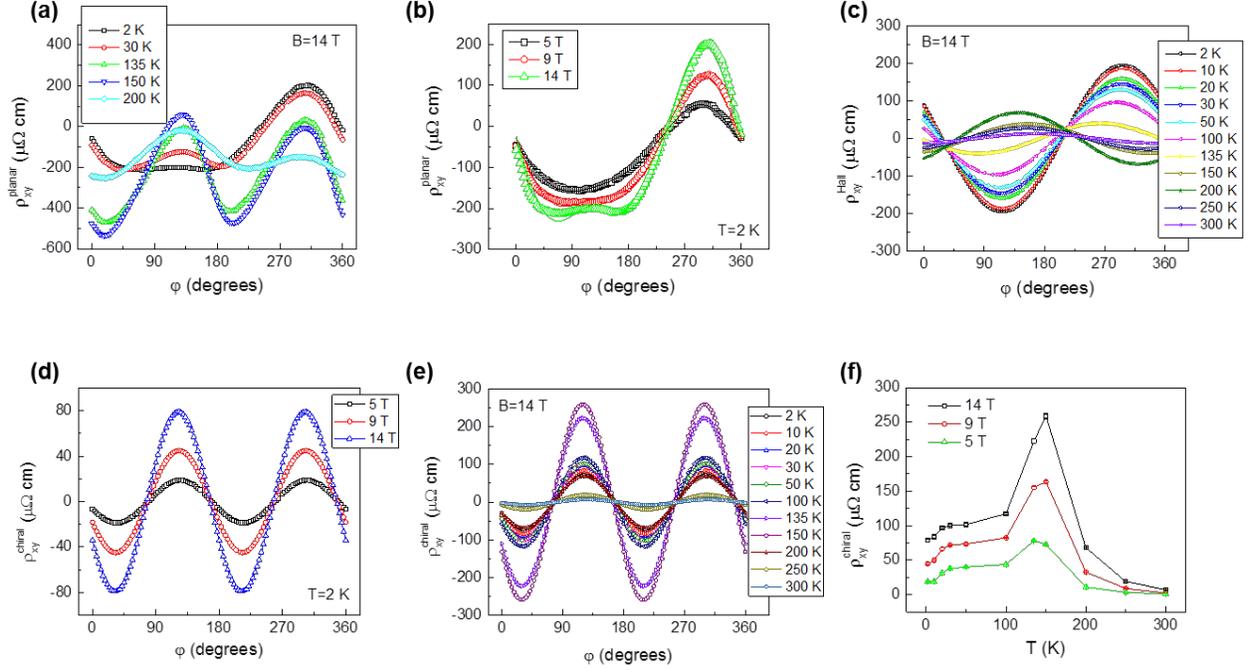

**FIG.9S** Another measurement of planar Hall effect on the same ZrTe$_{5-\delta}$ device (*t*=200 nm) in the main text. Ascribed to the different misalignment angle between the *ac* plane and magnetic field, The Hall effect component in planar Hall resistance is different from the data in FIG.3. (a) The angular dependence of planar Hall resistivity $\rho_{xy}^{planar}$ at different temperatures (*B*=14 T). (b) The angular dependent planar Hall resistance $\rho_{xy}^{planar}$ under different magnetic fields (*T*=2 K). (c) The extracted Hall resistance component $\rho_{xy}^{Hall}$ from $\rho_{xy}^{planar}$ at different temperatures (*B*=14 T). The 180° phase shift across ~135 K is ascribed to the change of dominated carrier type. This feature in the second set of data agrees well with that in main text. (d) The extracted intrinsic angular dependent planar Hall resistivity $\rho_{xy}^{chiral}$ caused by chiral anomaly under different magnetic fields (*T*=2 K). (e) The angular dependent intrinsic planar Hall resistivity $\rho_{xy}^{chiral}$ at different temperatures (*B*=14 T). (f) Temperature dependent intrinsic planar Hall resistivity shows a peak near 135 K, which is similar to the behavior in FIG. 4 in main text.

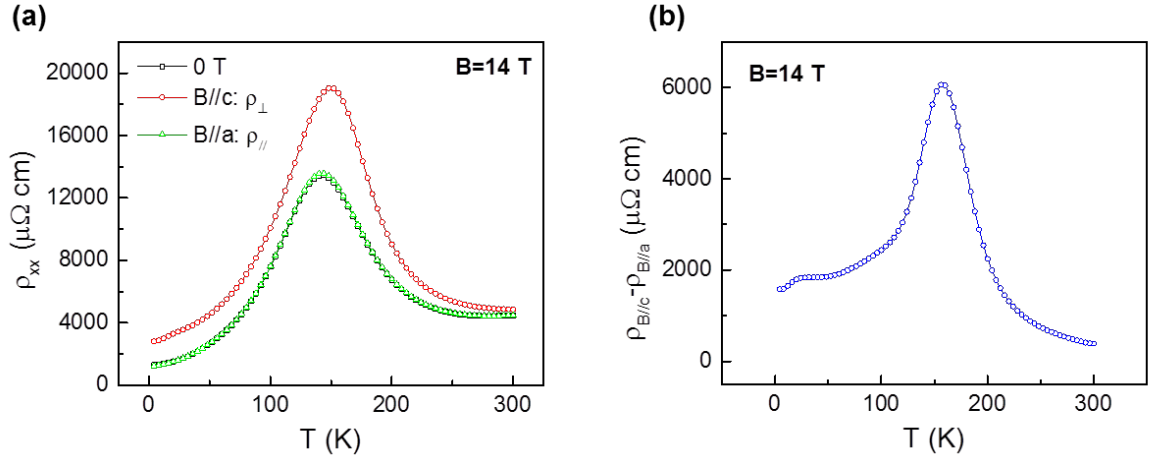

**FIG.10S** (a) Temperature dependence of ZrTe$_{5-\delta}$ resistivity ($t$=200 nm) with the magnetic field parallel to $c$-axis ($\rho_\perp$), $a$-axis ($\rho_{//}$), and the case without magnetic field ($B$=0). The temperature dependence of chiral-anomaly-induced resistivity anisotropy ($\Delta\rho_{\text{chiral}} = \rho_\perp - \rho_\parallel$) is calculated and given in (b). This temperature dependence is completely the same as that we observed in Fig. 4(c) and Fig. 9S(f). The chiral anomaly reaches its maximum near 150 K and persisted up to 300 K.

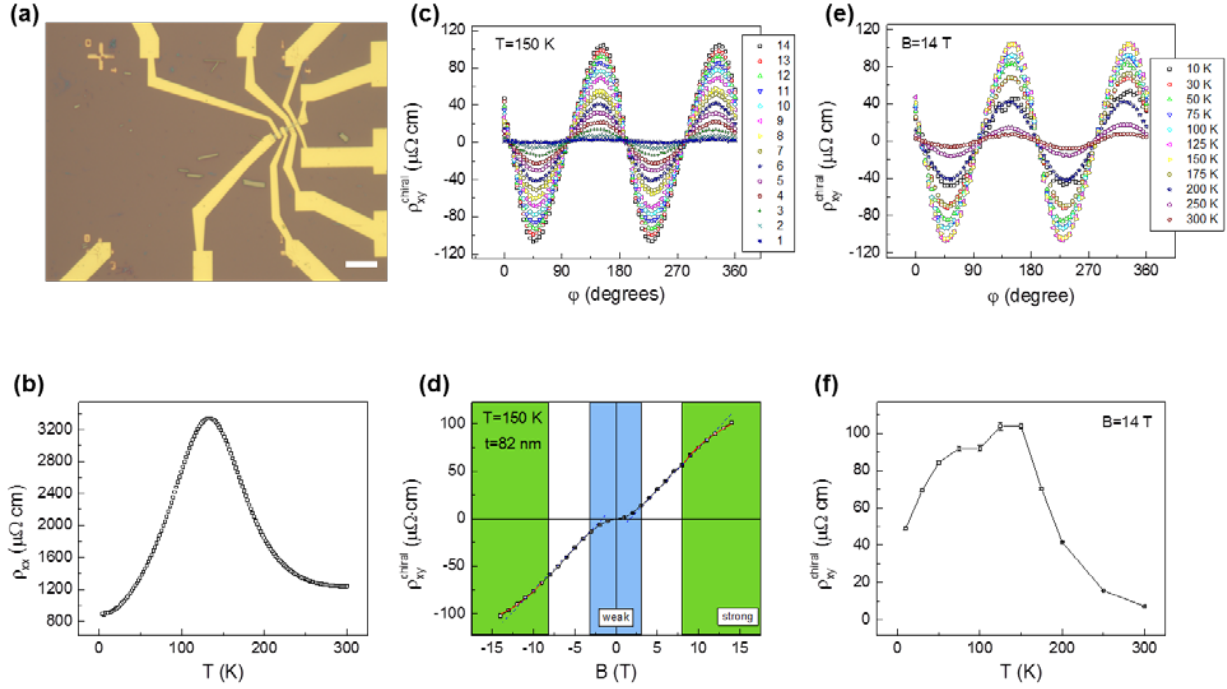

**FIG.11S** Reproducible planar Hall effect in ZrTe$_{5-\delta}$ devices ($t$=82 nm). (a) Optical image of ZrTe$_{5-\delta}$ devices, and the scale bar is 10 μm. (b) Temperature dependence of resistivity of ZrTe$_{5-\delta}$ device. (c) Angular dependence of intrinsic chiral-anomaly-induced planar Hall resistivity under different magnetic fields ($T$=150 K). (d) Field dependence of planar Hall resistivity ($T$=150 K). The weak tendency of saturation is distinguishable at high magnetic fields. (e) Angular dependence of planar Hall resistivity at different temperatures ($B$=14 T). (f) Temperature dependence of planar Hall resistivity under a magnetic field of 14 T.

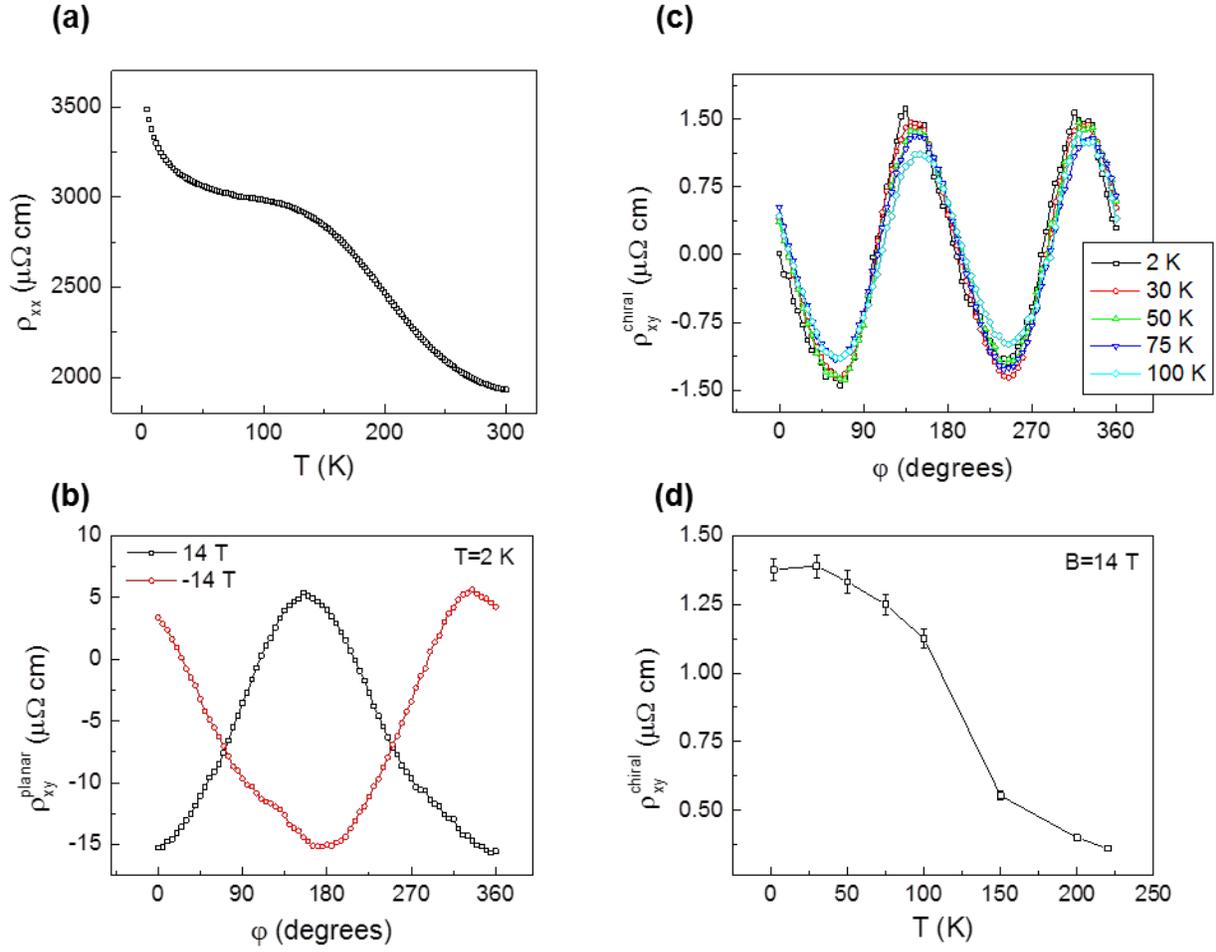

**FIG.12S** Planar Hall effect in thin $ZrTe_{5-\delta}$ device ($t$=30 nm). (a) Temperature dependence of resistivity. (b) Angular dependent planar Hall resistivity under a magnetic field of 14 T and -14 T ($T$=2 K). (c) Intrinsic planar Hall resistivity as a function of in-plane azimuthal angle. (d) Temperature dependent chiral-anomaly-induced planar Hall resistivity.

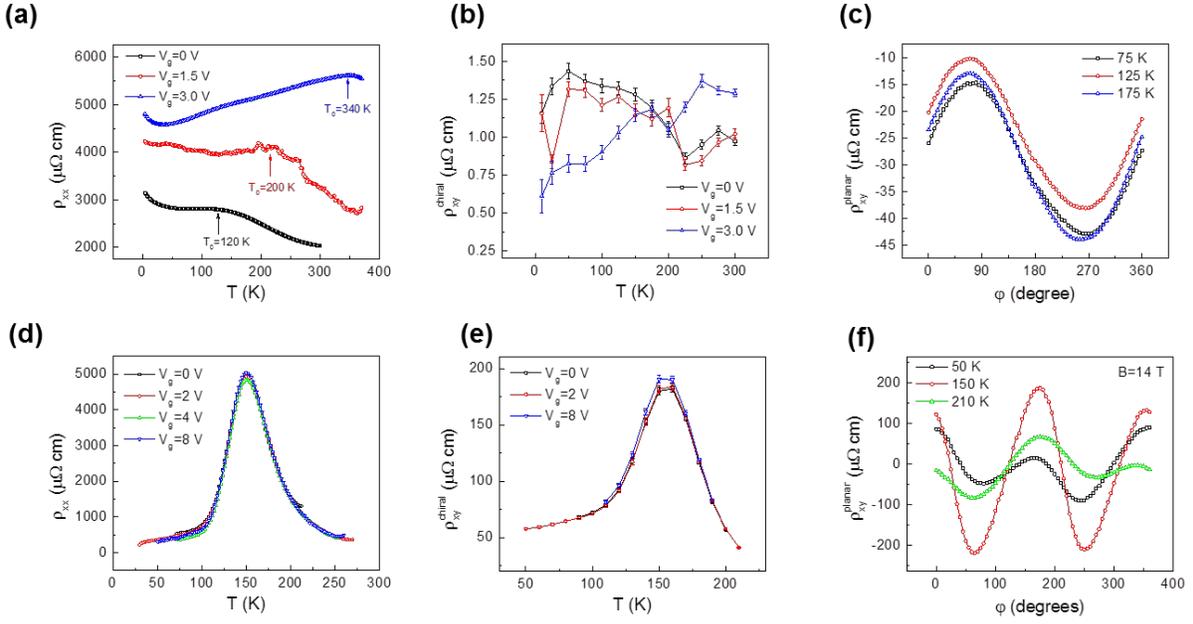

**FIG.13S** Gating effect on the planar Hall effect in ZrTe$_{5-\delta}$ devices. (a)-(c) *t*=25 nm; (d)-(f) *t*=174 nm. (a) and (d) represent the temperature dependent resistivity under different ionic gating voltages. (b) and (e) give the temperature dependent planar Hall resistivity under different gating voltages. (c) and (f) illustrate the typical angular dependent planar Hall resistivity near the transition temperature. The thick ZrTe$_{5-\delta}$ device (174 nm) displays much more distinct planar Hall effect than thin device (25 nm).

## Supplementary References